\documentclass[11pt, oneside]{amsart}
\pdfoutput=1

\usepackage{amsmath}
\usepackage{amssymb}

\usepackage{color}
\usepackage{dcolumn}
\usepackage{float}
\usepackage{graphicx}
\usepackage[latin9]{inputenc}
\usepackage{multirow}
\usepackage{rotating}
\usepackage{subfigure}
\usepackage{psfrag}
\usepackage{tabularx}
\usepackage[hyphens]{url}
\usepackage{wrapfig}
\usepackage{longtable}
\usepackage{verbatim}

\usepackage{footnote}
\makesavenoteenv{tabular}
\makesavenoteenv{table}

\usepackage{enumitem}
\setlist{leftmargin=7mm}

\usepackage[bookmarks, bookmarksopen, bookmarksnumbered]{hyperref}
\usepackage[all]{hypcap}
\definecolor{orange}{rgb}{1.0,0.3,0.0}
\definecolor{violet}{rgb}{0.75,0,1}
\definecolor{darkgreen}{rgb}{0,0.6,0}
\definecolor{cyan}{rgb}{0.2,0.7,0.7}
\definecolor{blueish}{rgb}{0.2,0.2,0.8}

\newcommand{\note}[1]{ {\textcolor{blueish}    { ***Note:      #1 }}}

\usepackage[top=2.5cm, bottom=2.5cm, left=2.5cm, right=2.5cm]{geometry}

\sloppypar

\begin{document}

\title[]{Report on the Second Workshop on Sustainable Software for Science: Practice and Experiences (WSSSPE2)}

\author{Daniel S. Katz$^{(1)}$, Sou-Cheng T. Choi$^{(2)}$, Nancy Wilkins-Diehr$^{(3)}$, Neil Chue Hong$^{(4)}$,
\\Colin C. Venters$^{(5)}$, James Howison$^{(6)}$, Frank Seinstra$^{(7)}$, Matthew Jones$^{(8)}$,
\\Karen Cranston$^{(9)}$, Thomas L. Clune$^{(10)}$, Miguel de Val-Borro$^{(11)}$, Richard Littauer$^{(12)}$}
\thanks{{}$^{(1)}$ Computation Institute, 
University of Chicago \& Argonne National Laboratory, Chicago, IL, USA; \url{dsk@uchicago.edu}}
\thanks{{}$^{(2)}$ NORC at the University of Chicago and Illinois Institute of Technology, Chicago, IL, USA; \url{sctchoi@uchicago.edu}}
\thanks{{}$^{(3)}$ University of California-San Diego, San Diego, CA, USA; \url{wilkinsn@sdsc.edu}}
\thanks{{}$^{(4)}$ Software Sustainability Institute, 
University of Edinburgh, Edinburgh, UK; \url{N.ChueHong@software.ac.uk}}
\thanks{{}$^{(5)}$ University of Huddersfield, School of Computing and Engineering, Huddersfield, UK; \url{C.Venters@hud.ac.uk}}
\thanks{{}$^{(6)}$ University of Texas at Austin, Austin, TX, USA; \url{jhowison@ischool.utexas.edu}}
\thanks{{}$^{(7)}$ Netherlands eScience Centre, Amsterdam, Netherlands; \url{F.Seinstra@esciencecenter.nl}}
\thanks{{}$^{(8)}$ National Center for Ecological Analysis and Synthesis, Santa Barbara, CA, USA; \url{jones@nceas.ucsb.edu}}
\thanks{{}$^{(9)}$ National Evolutionary Synthesis Center, Durham, NC, USA; \url{karen.cranston@nescent.org}}
\thanks{{}$^{(10)}$ NASA Goddard Space Flight Center, Greenbelt, MD, USA; \url{Thomas.L.Clune@nasa.gov}}
\thanks{{}$^{(11)}$ Department of Astrophysical Sciences, 
Princeton University, Princeton, NJ, USA; \url{mdevalbo@astro.princeton.edu}}
\thanks{{}$^{(12)}$ University of Saarland, Germany; \url{richard.littauer@gmail.com}}

\begin{abstract}
This technical report records and discusses the Second Workshop on Sustainable
Software for Science: Practice and Experiences (WSSSPE2). 
The report includes a description of the alternative, experimental submission
and review process, two workshop keynote presentations, a series of lightning
talks, a discussion on sustainability, and five discussions from the topic areas
of exploring sustainability; software development experiences; credit \&
incentives; reproducibility \& reuse \& sharing; and code testing \& code
review. For each topic, the report includes a list of tangible actions that were
proposed and that would lead to potential change.
The workshop recognized that reliance on scientific software
is pervasive in all areas of world-leading research today. The workshop
participants then proceeded to explore different perspectives on the concept of
sustainability. Key enablers and barriers of sustainable scientific software
were identified from their experiences. In addition,
recommendations with new requirements such as software credit files and software
prize frameworks were outlined for improving practices in sustainable software
engineering.
There was also broad consensus that formal
training in software development or engineering was rare among the
practitioners. Significant strides need to be made in building a sense of
community via training in software and technical practices, on increasing their
size and scope, and on better integrating them directly into graduate education
programs.
Finally, journals can define and publish policies to improve reproducibility, whereas
reviewers can insist that authors provide sufficient information and access to
data and software to allow them reproduce the results in the paper. Hence a list of
criteria is compiled for  journals to provide to reviewers so as to make it easier to
review software submitted for publication as a ``Software Paper.''

\end{abstract}

\maketitle
\newpage

\section{Introduction} \label{sec:intro}

%
%
%
%


The Second Workshop on Sustainable Software for Science: Practice and
Experiences
(WSSSPE2)\footnote{\url{http://wssspe.researchcomputing.org.uk/wssspe2/}} was
held on 16 November, 2014 in the city of New Orleans, Louisiana, USA, in
conjunction with the International Conference for High Performance Computing,
Networking, Storage and Analysis
(SC14)\footnote{\url{http://sc14.supercomputing.org}}. WSSSPE2 followed the
model of a general initial workshop,
WSSSPE1\footnote{\url{http://wssspe.researchcomputing.org.uk/wssspe1/}}~\cite{WSSSPE1-pre-report,WSSSPE1},
which co-occurred with SC13, and a focused workshop,
WSSSPE1.1\footnote{\url{http://wssspe.researchcomputing.org.uk/wssspe1-1/}},
which was organized in July 2014 jointly with the SciPy
conference\footnote{\url{https://conference.scipy.org/scipy2014/participate/wssspe/}}.

Progress in scientific research is dependent on the quality and accessibility of
software at all levels. Hence it is critical to address challenges related to
the development, deployment, maintenance, and overall sustainability of reusable
software as well as education around software practices. These challenges can be
technological, policy based, organizational, and educational, and are of
interest to developers (the software community), users (science disciplines),
software-engineering researchers, and researchers studying the conduct of
science (science of team science, science of organizations, science of science
and innovation policy, and social science communities). The WSSSPE1 workshop
engaged the broad scientific community to identify challenges and best practices
in areas of interest to creating sustainable scientific software. WSSSPE2
invited the community to propose and discuss specific mechanisms to move towards
an imagined future practice for software development and usage in science and
engineering. The workshop included multiple mechanisms for participation,
encouraged team building around solutions, and identified risky solutions with
potentially transformative outcomes. It strongly encouraged participation of
early-career scientists, postdoctoral researchers, and graduate students, with
funds provided to the conference organizers by the Moore Foundation and the
National Science Foundation (NSF), to support the travel of potential
participants who would not otherwise be able to attend, and young participants
and those from underrepresented groups, respectively. These funds allowed 17
additional participants to attend, and each was offered the chance to present a
lightning talk.

This report extends a previous report that discussed the submission,
peer-review, and peer-grouping processes in detail~\cite{WSSSPE2-pre-report}. It
is also based on a collaborative set of notes taken with Google Docs during the
workshop~\cite{WSSSPE2-google-notes}. Overall, the report discusses the
organization work done before the workshop (\S\ref{sec:preworkshop}); the
keynotes (\S\ref{sec:keynotes}); a series of lightning talks
(\S\ref{sec:lightning}), intended to give an opportunity for attendees to
quickly highlight an important issue or a potential solution; a session on
defining sustainability (\S\ref{sec:defining}). The report also gives summaries
of action plans proposed by five breakout sessions, which explored in specific
areas including sustainability (\S\ref{sec:exploring}); software development
experiences (\S\ref{sec:devel}); credit \& incentives (\S\ref{sec:credit});
reproducibility, reuse, \& sharing (\S\ref{sec:reproduce}); code testing \& code
review (\S\ref{sec:code_testing}). Lastly, the report also includes some
conclusions (\S\ref{sec:conclusions}) and an incomplete list of attendees
(Appendix~\ref{sec:attendees}).

\section{Submissions, Peer-Review, and Peer-Grouping} \label{sec:preworkshop}


WSSSPE2 began with a call for papers~\cite{WSSSPE2-pre-report}. Based on the
goal of encouraging a wide range of submissions from those involved in software
practice, ranging from initial thoughts and partial studies to mature
deployments, but focusing on papers that are intended to lead to changes, the
organizers wanted to make submission as easy as possible. The call for papers
stated:

\begin{quote} We invite short (4-page) \textbf{actionable} papers that will lead
to improvements for sustainable software science. These papers could be a call
to action, or could provide position or experience reports on sustainable
software activities. The papers will be used by the organizing committee to
design sessions that will be highly interactive and targeted towards
facilitating action. Submitted papers should be archived by a third-party
service that provides DOIs. We encourage submitters to license their papers
under a Creative Commons license that encourages sharing and remixing, as we
will combine ideas (with attribution) into the outcomes of the workshop.
\end{quote}

The call included the following areas of interest:
\begin{quote}
\begin{itemize}
\renewcommand{\labelenumi}{\textbf{\theenumi}.}
\setlength{\rightmargin}{1em}
\item defining software sustainability in the context of science and engineering
software
\begin{itemize}
\item how to evaluate software sustainability
\end{itemize}

\item improving the development process that leads to new software
\begin{itemize}
\item methods to develop sustainable software from the outset
\item effective approaches to reusable software created as a by-product of
research
\item impact of computer science research on the development of scientific
software
\end{itemize}

\item recommendations for the support and maintenance of existing software
\begin{itemize}
\item software engineering best practices
\item governance, business, and sustainability models
\item the role, operation, and
sustainability of community software repositories
\item reproducibility, transparency needs that may be unique to science
\end{itemize}

\item successful open source software implementations
\begin{itemize}
\item incentives for using and contributing to open source software
\item transitioning users into contributing developers
\end{itemize}

\item building large and engaged user communities
\begin{itemize}
\item developing strong advocates
\item measurement of usage and impact
\end{itemize}

\item encouraging industry's role in sustainability
\begin{itemize}
\item engagement of industry with volunteer communities
\item incentives for industry
\item incentives for community to contribute to industry-driven projects
\end{itemize}

\item recommending policy changes
\begin{itemize}
\item software credit, attribution, incentive, and reward
\item issues related to multiple organizations and multiple countries, such as
intellectual property, licensing, etc.
\item mechanisms and venues for publishing software, and the role of publishers
\end{itemize}

\item improving education and training
\begin{itemize}
\item best practices for providing graduate students and postdoctoral
researchers in domain communities with sufficient training in software
development
\item novel uses of sustainable software in education (K-20)
\item case studies from students on issues around software development in the
undergraduate or graduate curricula
\end{itemize}

\item careers and profession
\begin{itemize}
\item successful examples of career paths for developers
\item institutional changes to support sustainable software such as promotion
and tenure metrics, job categories, etc.
\end{itemize}

\end{itemize}

\end{quote}

31 submissions were received; all but one used arXiv\footnote{\url{http://arxiv.org}}
or figshare\footnote{\url{http://figshare.com}} to self-publish their papers.

The review process was fairly standard. First, reviewers bid for papers. Then an
automated system matched the bids to determine assignments. After the reviewers
completed their assigned reviews (with an average of 4.9 reviews per paper and
4.1 reviews per reviewer), they used EasyChair\footnote{\url{http://easychair.org/}} 
to record scores on relevance
and comments. Finally, the organizers accessed the information to decide which
papers to associate with the workshop and provided authors with the comments to
help them improve their papers.

The organizers decided to list 28 of the papers as significantly contributing to
the workshop, a very high acceptance rate, but one that is reasonable, given the
goal of broad participation and the fact that the reports were already
self-published.

The organizers wanted very interactive sessions, with the process of creating
the sessions open to the full program committee, the paper authors, and others
who might attend the workshop. In order to do this, the organizers used 
Well Sorted\footnote{\url{http://www.well-sorted.org}} with the following steps:
\begin{enumerate}
\item Authors were asked to create Well Sorted ``cards'' for the papers. These
cards have a title (50 characters maximum) and a body (255 characters maximum).
\item Authors, members of the WSSSPE program committee, and mailing list subscribers
were asked to sort the cards. Each person dragged the cards, one by one, into
groups. A group could have as many cards as the person wanted it to have, and it
could have any meaning that made sense to that person.
\item Well Sorted produced a set of averages of all the sorts, with
various numbers of card clusters.
\end{enumerate}

The organizers then chose a sort that contained five groups that felt most
meaningful. After that, they decided on names for the five groups:
\begin{itemize}
\item Exploring Sustainability
\item Software Development Experiences
\item Credit \& Incentives
\item Reproducibility \& Reuse \& Sharing
\item Code Testing \& Code Review.
\end{itemize}

Finally, since some of the papers were not represented by cards in the process,
they were not placed in groups by the peer-grouping system. The authors of
these papers were asked which groups seemed the best for their papers; these
papers were then placed in those groups. Sections~\ref{sec:exploring}-\ref{sec:code_testing}
discuss the breakout groups, including a list of the papers associated with each
group.

\section{Keynotes} \label{sec:keynotes}

The workshop featured two keynote addresses. In the opening keynote
presentation, Kaitlin Thaney of the Mozilla Science Lab talked about her
organization's work and policy to enable and support sustainable and
reproducible scientific research through the open web. The second keynote
speaker was Neil Chue Hong of Software Sustainability Institute. He shed
light on how scientific software is prevalently driving advances in many science
and engineering fields. Both keynote speeches spawned further discussion among
workshop participants on the crucial notion of \emph{software sustainability}
in the theme of our workshop.

\subsection{Kaitlin Thaney, Designing for Truth, Scale, and Sustainability~\cite{Thaney_slides}}
\label{keynote1}

Kaitlin Thaney is the Director of the Mozilla Science Lab (hereafter Mozilla), which
is a non-profit organization interested in openness, news, website
creation, and Science, all taking advantage of the open web.

Thaney started noting the unfortunate fact that many current systems suffer the
unintended consequence of creating friction that hinders users, despite
designers' original purposes to do good. An example is the National Cancer
Institute's caBIG. A total of \$350 million was spent, including more than \$60
million for management. More than $70$ tools were created, but caBIG is still
seen as a failure\footnote{Report Blasts Problem-Plagued Cancer Research Grid,
\url{http://tinyurl.com/maf6dz2}}. Those that had the least investment were the
most used; the most invested software were the least utilized.

Thaney emphasized that for efficient reproducible open research, we would need
research tools (e.g., software repositories), social capital (e.g., incentives),
and capacity (e.g., training and mentorship). Our systems would need to
communicate with each other. A point was made by a member of the audience that
as systems become less monolithic, it often becomes harder to sustain the links
between them\footnote{See, for example, \url{http://tinyurl.com/l76tba2}.}.

Thaney spoke about Mozilla's work around code citation, through a collaboration
and prototype crafted between Mozilla, GitHub, figshare, and Zenodo. This work
was presented at a closed meeting in May 2014 at the National Institutes of
Health (NIH) around these issues, sparking a conversation from that meeting
around what a \emph{Software Discovery
Index}\footnote{Software Discovery Index, \url{http://softwarediscoveryindex.org/report/}} might look
like. The meeting included a number of publishers, researchers, and those behind
major scientific software efforts such as
Bioconductor\footnote{Bioconductor, \url{http://www.bioconductor.org}},
Galaxy\footnote{Galaxy, \url{http://galaxyproject.org}}, and
nanoHUB\footnote{nanoHUB, \url{https://nanohub.org}}.
Ted Habermann in the audience commented that if the metadata is minimal, it
would be less onerous for data providers, but more burdensome for users---it
could be challenging to keep a balance between what have to be captured and what
would be ideal if we do not want to lose user engagement as in the case of the
old Harvard Dataverse, finding often only the first four fields of three pages
of metadata were filled out.

The speaker concluded her talk urging the audience to design scientific software
with the general community, not an individual, in mind; and to design to unlock
latent potential of our systems. In addition, she encouraged everyone to rethink
how we reward researchers and support roles.

\subsection{Neil Chue Hong, We are the 92$\%$~\cite{Hong_slides}}
\label{keynote2}

Neil Chue Hong is the Director of the Software Sustainability Institute (SSI) in the
United Kingdom (UK). The SSI was founded to support the UK's research software
community by cultivating better, more sustainable research software to enable
world-class research. Chue Hong's keynote started by making the point
that the use of -- and reliance on -- software is pervasive in all areas of
world-leading research, showing examples from disciplines as diverse as
humanities and high-energy physics, quoting Kerstin Kleese Van Dam of the
Pacific Northwestern National Laboratory via a petition
campaign\footnote{\url{http://tinyurl.com/nkn5tnv}}
at change.org, ``\emph{Today there are very few science areas left which do not
rely on IT and thus software for the majority of their research work. More
importantly key scientific advances in experimental and observational science
would have been impossible without better software.}'' He also cited Daniel
Katz, Software Infrastructure for Sustained Innovation (SI2) Program Director of
the NSF, ``\emph{Scientific discovery and innovation are advancing along
fundamentally new pathways opened by development of increasingly sophisticated
software. Software is an integral enabler of computation, experiment and theory,
and directly responsible for increased scientific productivity and enhancement
of researchers' capabilities.}''

Chue Hong drew attention to the issue that in the cyber-infrastructure and
high-performance community, hundreds of thousands
of researchers developing software are all too often disregarded or considered the long tail.
Actually, the numbers point to the fact that they are the mainstream. 
He emphasized that software is no longer special; it is both essential to and
common in scientific research. A 2014
survey\footnote{\url{http://tinyurl.com/ooajs7m}}
conducted by the SSI polled researchers from 15 research-intensive UK
universities (406 respondents covering a representative range of funders,
disciplines, and seniority). The survey reported that 92\% respondents confirmed the use of research
software and 89\% affirmed that it would be impossible or difficult to conduct
research without software. Nevertheless the British research community is just
starting to understand the magnitude of the issue. Whilst many researchers make
use of software such as MATLAB, SPSS, and Excel, data from the aforementioned
SSI survey shows that over half (56\%) of respondents developed their own
research software (which equates to over 140,000 researchers if extrapolated
across the UK) and yet 71\% of all UK researchers had no formal software-development training, having
to rely on their own coding skills.

Examining another aspect of the size of the research software community,
Chue Hong noted that the costs of software-reliant research in the UK
included \textsterling 840 million of investment in the financial year
2013--2014, and this amount has risen by 3\% on average over the past four
years. About 30\% of total research investment has been spent on research that
relies on software over the last four financial years. These numbers stemmed
from an analysis by the SSI of data from $49,650$ grant titles and abstracts published on
Gateway to Research between years 2010 and 2014. A similar analysis of university
jobs advertised in the same period discovered that despite this investment, only
4\% of positions were software development related, and of these only 17\% were
explicitly named as a software developer or software engineer positions: the vast
majority being advertised as research associate or research assistant positions.
This in turn leads to the issue of career paths for those bridging the research
and software worlds, who are essential to support the use and further
development of research software, a point highlighted by a graphic showing UK
STEM graduate career paths\footnote{Source: The Scientific Century, Royal
Society, 2010 (revised to reflect first stage clarification from ``What Do PhD's
Do?'' study)} showing that only 3.5\% were able to secure permanent positions.

To conclude, Chue Hong led the audience in discussing the following questions:
What are \emph{we} going to \emph{do} to help and benefit the 92\% of researchers
who rely on software? Who do we need to persuade? What are the incentives we
need to put in place? Finally, he challenged the workshop participants to change
the current deficient practices in research and academia.

\section{Lightning Talks} \label{sec:lightning}

Lightning talks were a new feature in WSSSPE2. Since the workshop program was
mostly dedicated to discussions, the organizers wanted to give the attendees a
chance to also `make a pitch' for an idea, either representing a contributed
paper or something different. Eighteen attendees volunteered to participate in
the lightning talks, each given only two minutes to speak and at most one slide
to show. The talks were presented in reverse alphabetic order of speakers' last
names. In the rest of this section, we highlight the gist of some of the
speakers' messages.


\begin{enumerate}
\item \textbf{Colin C. Venters: The Nebuchadnezzar Effect: Dreaming of Sustainable
Software through Sustainable Software Architectures~\cite{Venters_poster}.}
Venters proposed that sustainable software is a composite, first-class,
non-functional requirement (NFR) that is at a minimum a measure of a system's
maintainability and extensibility, but may also include other NFRs such as
efficiency (e.g., energy, cost), interoperability, portability, reusability,
scalability, and usability. To achieve technically sustainable software, Venters
suggested that software architectures are fundamental as they are the primary
carrier of system NFRs, i.e., pre-system understanding; and influence how
developers are able to understand, analyze, extend, test, and maintain a
software system, i.e., post-deployment system understanding. In addition,
Venters highlighted that sustainability of software architectures needs to be
addressed to endure different types of change and evolution in order to mitigate
architectural drift, erosion, and knowledge vaporization.



\item \textbf{Marlon Pierce: Patching It Up, Pulling It
Forward~\cite{Pierce_poster}.} Pierce discussed how open open source is. Open
software needs a diverse, openly governed community behind it, just as it needs
open licensing and a public code repository. To probe the level of governance
within open source projects, he and his co-authors (Marru and Mattmann)
suggested a contest to encourage individual developers to submit
patches and requests to projects that are important to them. This simple
mechanism shall expose several governance mechanisms, such as how easy it is for
independent developers to communicate with project leadership, how projects
accept and license third-party contributions, and how projects make
decisions such as granting source tree write access.

\item \textbf{John Peterson: Continuous Integration for Concurrent MOOSE Framework
and Application Development on GitHub~\cite{Peterson_poster}.}
Peterson from the Idaho National Laboratory reported
that in March 2014, the MOOSE framework was released under an open source
license on GitHub, significantly expanding and diversifying the pool of current
active and potential future contributors on the project. The MOOSE team employs
an extensive continuous integration test suite to ensure that both the framework
and the applications based on the framework are verified before any code changes
are accepted into the repository. They use a combination of built-in Git
features such as branching and submodules, GitHub API integration capabilities,
and in-house developed testing software to perform this verification and update
the dependent applications in a relatively seamless manner for users.

\item \textbf{Abani Patra: Execute it~\cite{Patra_poster}.}
Patra discussed the value of an easily accessible platform for
executing scientific software, e.g., HUBzero to access XSEDE or other computing
resources. Such a platform for executing benchmark problems (even at a small
scale) allows the developer community access a reference implementation and
provides an easy way to train the larger user community. A second idea of this
talk was that for true usability, much more attention and support needs
to be given to the integrated use of simulation tools inside complex workflows.


\item \textbf{Daniel S. Katz: Implementing Transitive Credit with
JSON-LD~\cite{Katz_transitive_credit_poster}.} Science and engineering research
increasingly relies on activities that facilitate research but are not rewarded
or recognized, such as: data sharing; developing common data resources, software
and methodologies; and annotating data and publications. To promote and advance
these activities, we must develop mechanisms for assigning credit, facilitate
the appropriate attribution of research outcomes, devise incentives for
activities that facilitate research, and allocate funds to maximize return on
investment. Katz discussed the issue of assigning credit for both direct and
indirect contributions by using JSON-LD to implement a prototype transitive
credit system.

\item \textbf{Samin Ishtiaq: Daemons, Notifications and Sustaining Software. }
The reproduction and replication of novel results has become a major issue in
computer science, systems biology, and other computational disciplines. These
include both the inability to re-implement novel algorithms and approaches, and
lack of an agreement on how and what to benchmark these algorithms on. Ishtiaq
from Microsoft Research Cambridge pointed out these problems and made several
suggestions to address them.

\item \textbf{James Howison: Retract all Bit-Rotten Publications. }
Howison sought to provoke discussion by proposing that papers whose workflows
are not kept current with the changing software ecosystem should be
automatically retracted. This would create an incentive for authors to keep
their software current and usable, rather than the current situation in which
every potential user has to do this individually. A softer version of the
proposal would identify papers whose software workflow has become bit-rotten and
allow others to keep the code up to date, either adding them as new authors of
the paper or providing credit for their academic service in some other form.




\item \textbf{Robert Downs: Community Recommendations for Improving Sustainable
Scientific Software Practices~\cite{Downs_poster}.} Robert Downs, of the
Columbia University Center for International Earth Science Information Network
(CIESIN), described a focus group study conducted with the Science Software
Cluster (SSC) of the Federation of Earth Science Information Partners (ESIP).
For the study, almost 300 attendees of the 2014 Summer ESIP Meeting were invited
to participate in simultaneous roundtable discussions on sustainability of
science software. Over two-thirds of the roundtable focus groups responded to a
semi-structured survey that contained three sets of questions eliciting
recommendations for near-term actions of the community to improve sustainable
software practices. Initial analysis of the participants' responses to the
questionnaire revealed several suggestions, which included improving community
engagement and collaborative activities, increasing understanding and awareness,
and creating incentives to motivate sustainable science software practices. 
The ESIP SSC plans to engage the community in the recommended activities for
improving sustainable scientific software practices.

\item \textbf{Carl Boettiger: rOpenSci: Building Sustainable Software by
Fostering a Diverse Community~\cite{Boettiger_poster}.} Boettiger described how
the rOpenSci project has been successful by focusing not just on building
software but also on building a community of researchers who learn and adopt
their approaches to reproducible research and sustainable software practice.
Through outreach, mentoring, workshops, and hackathons, they have not only
reached new users, but also turned users into co-developers of robust software
and good practices to support data science research across a growing set of
disciplines.

\item \textbf{Jakob Blomer: The Need for a Versioned Data Analysis
Environment~\cite{Blomer_poster}.} Large-scale scientific endeavors, such as the
discovery of the Higgs boson at the Large Hadron Collider (LHC), often rely on
complex software stacks. Maintaining thousands of dependencies of software
libraries and operating system versions has shown that despite source code
availability, the setup and the validation of a minimal usable analysis
environment can easily become prohibitively expensive. In high-energy physics,
CernVM-FS, a special-purpose, open-source, versioning, and snapshotting file
system used to capture and distribute entire software stacks, proved to be
useful for providing instant access to ready-to-run data analysis environments.

\item \textbf{Alice Allen: Find it! Cite it! }
The Astrophysics Source Code Library (ASCL) is an online registry of
scientist-written software used in astronomy research. Their primary interest is
rendering research more transparent by making this software more discoverable for
examination. The ASCL is treated as a publication by an indexing resource for
astronomy, the Astrophysics Data System (ADS). ADS tracks citations to what it
indexes, including citations to ASCL entries. Increasing rewards for writing
software, whether through citation, transitive credit or other methods, gives
software authors a powerful reason to take the time to build sustainability into
their software and is an excellent way to drive community change.
\end{enumerate}

\smallskip

\section{Defining Sustainability} \label{sec:defining}


In the first interactive session, the attendees divided themselves into groups
to discuss software sustainability. They were asked to
\begin{enumerate}
\item discuss what the term ``software sustainability'' meant to them

\item determine three things they considered to be significant enablers of
software sustainability

\item determine three things they considered to be significant barriers to
software sustainability.
\end{enumerate}
Once each group had come up with answers, all the answers were compiled, and the
attendees voted on which they thought were important by a show of hands.

The general responses to what software sustainability meant were:
\begin{itemize}
\item keeping software scientifically useful
\item separating techniques in code from knowledge in code
\item that an adequately large community finds value in software and is willing
to sustain it.
\end{itemize}

The enablers of and barriers to software sustainability, roughly ranked by
attendee votes are shown in
{\tablename}s~\ref{tb:software_sustainability_enablers} and
\ref{tb:software_sustainability_barriers}, respectively.\footnote{A few other
items were suggested as barriers, but were not voted on due to lack of time in
the session:
layering up dependencies;
using software past its sustainable life;
using software past its usable life;
inertia for accepted answers versus wrong or right answers;
monolithic or poor code; and
need to restructure code when hardware/software/libraries change.
}

\begin{table*}[ht]
\centering
\caption{Enablers of software sustainability, with 0 to 10 `*'s roughly
indicating the fraction of attendees who voted for an item as important.}
\label{tb:software_sustainability_enablers}
  \begin{small}
  \begin{tabular}{ | p{1.65cm} | p{12.0cm} |}
    \hline
  Importance & Item \\ \hline \hline
********** & healthy and vibrant communities; vibrant community to champion software \\ \hline
********** & designing for growth and extension -- open development \\ \hline
******* & culture in community for reuse \\ \hline
**** & portability \\ \hline
**** & culture in development community to support transition between developers \\ \hline
*** & interdisciplinary people: science + IT experience \\ \hline
** & planning for end of life \\ \hline
** & make smart choices about dependencies \\ \hline
* & thinking of software as product lines -- long term vs.~short term view \\ \hline
 & not all communities need new software \\ \hline
 & converting use into resources \\ \hline
    \end{tabular}
    \end{small}
\end{table*}

\begin{table}[ht]
\caption{Barriers to software sustainability, with 0 to 10 `*'s roughly
indicating the fraction of attendees who voted for an item as important.}
\label{tb:software_sustainability_barriers}
 \centering
  \begin{small}
  \begin{tabular}{ | p{1.65cm} | p{12.0cm} |}
   \hline
  Importance & Item \\ \hline \hline
******* & lack of incentives, including promotion and tenure process; promotion
and tenure process in academic is incompatible with sustainability \\ \hline
*****  & absent or poor documentation \\ \hline
***** & funding to ensure sustainability is difficult to obtain \\ \hline
**** & developers are not computer scientists; don't have software engineering
practices (in particular, those needed to scale-up projects to support and be
developed by a large sustainable community) \\ \hline
*** & overreliance on one or two people -- the `bus test'\footnote{``Bus test''
in Table~\ref{tb:software_sustainability_barriers} refers to the smallest number
of key people a project would need to lose to become non-viable (the larger the
number, the healthier the project).} \\ \hline
** & rate of change of underlying technologies \\ \hline
** & lack of business models for sustainability \\ \hline
* & lack of training for how to build sustainability into the system \\ \hline
 & maintenance for software is not visible, appears to ``just happen'' \\ \hline
 & licensing issues \\ \hline
 & staff turnover -- lack of continuity \\ \hline
\end{tabular}
 \end{small}
\end{table}

\section{Exploring Sustainability} \label{sec:exploring}


Six papers were categorized under the theme of Exploring Sustainability. The
group included four of the authors from the six papers submitted and an
additional number of participants who expressed an interest in the theme at the
workshop. Each paper had a different perspective on the concept of
sustainability, which ranged from the sustainment of communities to defining
sustainability as a first-class, composite non-functional requirement.

\subsection{Discussion and Actions}
\subsubsection{Discussion}

Each author was invited to outline the key action from their paper as a
potential discussion point for the group; where the authors were not present,
the group facilitators outlined the actionable points from their papers. The key
actions from the six papers included recommendations for improving practice in
sustainable software engineering~\cite{wssspe2_rosada_de_souza, wssspe2_downs};
development of \emph{Software and Infrastructure as a Service} as a mechanism for
fostering sustainable science communities~\cite{wssspe2_patra}; developer
incentives for code contributions to open source projects~\cite{wssspe2_pierce};
establishment of a set of software engineering principles based on scalability,
reproducibility, and energy efficiency~\cite{wssspe2_shi}; and applying software
architectures as a mechanism for architectural-level reasoning about
sustainability~\cite{wssspe2_venters}. The group took the position of viewing
sustainability from the perspective of addressing the challenges related to the
development, deployment, and maintenance of reusable
software~\cite{WSSSPE1}. 

The principal focus of the initial discussion
considered how to foster cultural change towards developing sustainable software
in academic environments. It was suggested that a new requirement driven by the
agencies funding research projects where software was an intrinsic part of
enabling the research program would be an additional element within the grant
proposal to provide a sustainability plan for sustaining the software. This type
of initiative would provide the necessary incentive and motivation for
researchers to consider how to sustain their software beyond the lifetime of the
project.

The discussion then focused on a need for a common language and a definition on
the concept of sustainability that moved beyond the current fuzzy definitions
where time was the simple measurement of sustainment. It was suggested that
there was a need to identify tangible actions that underpin sustainability that
developers could incorporate into the development stream of their software. This
prompted a debate regarding whether sustainability should be considered as a
non-functional requirement or software quality as defined within ISO/IEC
25010~\cite{iso25010}. The focus of the discussion was based on the paper by
Venters et al.~\cite{wssspe2_venters}, which suggested that sustainability is a
first-class, composite, non-functional requirement composed of a number of
sub-characteristics. It was generally agreed that maintainability and
extensibility were key qualities underlying sustainability. In addition, the
group also discussed what other non-functional requirements would contribute to
the development of sustainable software, e.g., reusability and scalability.
However, it was recognized that there was a need to identify appropriate metrics
and measures.

The group also discussed whether the concept of sustainability itself was a
barrier to achieving sustainable software. It was suggested that ``the first
rule of software sustainability is do not talk about software
sustainability.'' Instead there should be a focus on best software engineering
practice. Playing devil's advocate, it was asked that if the focus on the
concept of sustainability was ignored, what current software engineering
practices and principles could be utilized by software developers and domain
scientists to achieve sustainability? This raised the question of why existing
software engineering knowledge, such as that contained in
SWEBOK~\cite{Bourque2014}, was largely ignored and to what extent the
environment has a strong influence on practice within the scientific and
engineering community. As a result, there is a need to identify best practices
and reach a consensus beyond the theories that underpin the discipline of
software engineering. Similarly, how could we translate or distill some of the
key building blocks that underpin software engineering practice?

A final key point of the discussion was the role that software design and
patterns play before committing to writing a line of code. It was suggested that
modeling must play a major role in attaining sustainable software. The point was
raised that design involves making decisions and a mechanism would be necessary
for capturing these decision points. This introduced the idea of software
provenance that moved beyond commits in software repositories to how to capture
and maintain relationships between sources and design decisions to prevent
knowledge vaporization. Whether this could be achieved through software
architectures is an open-research challenge.

\subsubsection{Actions}

The main action to come from the group was a proposal to identify the ten best
software engineering practices similar to Philip Bourne's approach of Ten Simple
Rules~\cite{bourne-ten-simple-rules}.

\subsection{Papers}
The papers that were discussed in the Exploring Sustainability group are:
\begin{itemize}
\item Mario Rosado de Souza, Robert Haines, and Caroline Jay. Defining
sustainability through Developers' Eyes: Recommendations from an Interview
Study~\cite{wssspe2_rosada_de_souza}

\item Robert Downs, W. Christopher Lenhardt, Erin Robinson, Ethan Davis, and
Nicholas Weber. Community Recommendations for Sustainable Scientific
Software~\cite{wssspe2_downs}

\item Abani Patra, Matthew Jones, Steven Gallo, Kyle Marcus, and Tevfik Kosar.
Role of Online Platforms, Communications and Workflows in Developing Sustainable
Software for Science Communities~\cite{wssspe2_patra}

\item Marlon Pierce, Suresh Marru, and Chris Mattmann. {WSSSPE2}: Patching It Up,
Pulling It Forward~\cite{wssspe2_pierce}

\item Justin Shi. Seeking the principles of sustainable software
engineering~\cite{wssspe2_shi}

\item Colin C. Venters, Michael K. Griffiths, Violeta Holmes, Rupert R. Ward, and
David J. Cooke. The Nebuchadnezzar Effect: Dreaming of Sustainable Software
through Sustainable Software Architectures~\cite{wssspe2_venters}
\end{itemize}

\section{Software Development Experiences} \label{sec:devel}


Of the short actionable papers that would lead to improvements for sustainable
software science, 11 submissions were categorized in the Software Development
Experiences group. Because of the large number, we split these into two
subgroups prior to the event. Some common themes helped in this division. For
example, several papers that addressed education and training issues, including
best practices and case studies, were grouped together. Others discussed
experiences with registries, developer collectives and specific examples of
successful, sustainable software (including a valuable industry perspective).

\subsection{Discussion and Actions}

Subgroup A consisted of ten participants who discussed papers surrounding training
and successful community software initiatives.
Subgroup B had six participants who discussed four papers. Because of the nature
of the papers, training emerged as a common theme. However the conversation was
wide-ranging, including incentives, reproducibility, and funding to promote
sustainability.  In the end, both groups discussed training, though from somewhat different
points of view and resulting in somewhat different suggested actions.

\subsubsection{Subgroup A Discussion}


Group discussion started with a position statement by each person surrounding
what they had learned from their software development experiences and how those
lessons might be translated into actionable outcomes. Participants came from a
range of backgrounds, and represented multiple software and training
initiatives, including Software Carpentry, Data Carpentry, Open Science for
Synthesis (OSS), the Community Surface Dynamics Modeling Systems group, ROpenSci,
DataONE, the HDF Group, and others. Some of these software experiences were
focused on development of new products for use in the sciences (e.g., ROpenSci,
CSDMS), and these recounted the difficulties of engaging with disciplinary
scientists in writing software. Software was clearly utilized broadly across the
various science disciplines represented, and it was developed within those
disciplines as well. Some researchers created software for statistical analysis
and modeling, while others used it to control instrumentation, collate data
across networks, collect information from respondents, and many other uses.
There was broad consensus that, within the disciplines represented, formal
training in any type of software development or engineering was rare among the
practitioners; most are self-taught, and develop software to get another job
done. Any ancillary utility of the software outside of the specific science
target was generally unplanned and few researchers would want to invest more
time to make their own software more re-usable.

There was general agreement that this body of disciplinary software improvement
needed to be understood in terms of the scientific productivity that could be
achieved. A software maturity model is needed for science software, but it needs
to be introduced in a way that fits the culture of science, which largely thinks
of software as a tool, rather than a product itself. The group was in general
consensus that more widespread training in software practice is needed within
the domain sciences, and several of the participants were involved in efforts
along these lines. Participants felt that projects that build a sense of
community via training in software and technical practices would have the most
success in changing practices in that community, but that there was a need for a
managed introduction of these practices. Participants also recognized that these
could not be one-time, one-off training opportunities, as software and
technologies change rapidly over time. For example, while today Software
Carpentry is focused on teaching version control via Git, there has been a rapid
evolution from RCS to CVS to SVN to Git over a short time frame, and thus
communities should expect the need to train for adaptability and a changing
technology tool chain, rather than assume that these technologies will stay
fixed. Thus, although short term training that introduces specific tools was
considered highly valuable, these trainings were also not considered sufficient
to engender the changes in software practice that were deemed necessary.
Combining the need for changes in practice to be introduced incrementally with
the need to minimize the divergence of training from direct science goals and
the difficulties of training for a rapidly evolving technology space,
participants concluded that multiple training efforts that targeted different
parts of the spectrum were needed. Short-term courses introducing immediately
useful skills needs to be offered alongside more in-depth courses on software
engineering and practice that allow students to adapt to a changing landscape.

Finally, after agreeing that these complementary training models were needed,
the group discussed sustainability of training, and how the leading groups in
this space are teaching only a small fraction of the community that needs and
wants training. Most graduate programs in the sciences do not currently
incorporate these approaches in their graduate curricula, although there is an
increasing number of quantitatively oriented courses around analysis and
modeling. These still, however, generally omit engineering practices such as
version control, unit and regression testing, and software modularity and
abstraction, often because the instructors themselves in the domain sciences are
not familiar with these techniques. Thus, students who are being trained in
these approaches are doing so through short 2--3 day training workshops such as
Software and Data Carpentry, rather than through semester-long graduate
education courses at their universities, which tend to focus on statistical and
modeling techniques. Hybrid programs like the three-week Open Science for
Synthesis (OSS) training that combine the three (science, quantitative
techniques, and software engineering) into an holistic course serve part of the
need but reach only a few researchers at this time. Thus, the group concluded
that significant strides needed to be made in coordinating these trainings so
that they are complementary, on increasing their size and scope, and on better
integrating them directly into graduate education programs.

\subsubsection{Subgroup B Discussion}
The statement ``applied computer science is being attempted in academia without
any formal training'' kicked off our discussion. The group brought expertise in
several different training models, from a two-day Software Carpentry workshop,
a three-week Open Science for Synthesis (OSS) program, to semester-long
programs. We discussed the pros and cons of different training approaches,
touching on informal learning, for example, where people learn the necessary
skills by asking questions of cross-disciplinary people (``boundary scientists'') in
their work environments.

Some papers explored gaps in training of early-career scientists. Industry
participants in our group confirmed this observation. We asked ourselves, ``Are traditional
courses failing?'' We think yes. Changes to undergraduate curricula requirements
are difficult. But as programming models become more complex, we have to raise
the skill levels. Skills must be improved not only in traditional programming --
learning languages and algorithms -- but also around professional software
development. We want to get to a point where ``Everyone has a new minimum now --
everyone knows Git.'' Developers also need training on licensing choices. Just
because some source code is on Git does not mean that it is open.

``How do you get people to look for the training they need?'' wondered one
participant. People seek out opportunities like Software Carpentry to augment
skills. While some instructions can be done in institutional curricula,
independent groups (non-profits, institutes) have more flexibility. Some asked
whether Silicon Valley would be interested in funding training so that people are
better prepared to enter the workforce. Some felt that companies were reluctant
to deliver training for fear that their employees might then leave them. While others felt
that pushing all training to industry could lead to good technical people
``getting on the Google bus'' and leaving sciences.

``How will we know when people are trained effectively in these new skills?'' We
discussed certification. It can be hard to build the recognition of Java or
database certifications among all technologies. OSS offers badges to those
completing training. However we need to demonstrate proficiency, not just
completion. Google Summer of Code is a big CV augmentation for
participants---can we create something similar?

The group also discussed how to fund training. On participant observed that NYU
runs a six-week data-science training; companies grab the graduates and pay for
those they hire. OSS also used the NSF Software Institutes program as a vehicle
to fund training activities. Software Carpentry uses a collaborative teaching
approach where people publish open teaching materials and receive credits for
their use.

We then discussed career paths for those supporting sustainable software. While
tenure track is not the only option for graduate students, the challenges for those who
remain in academia can be large. Research scientists are entirely dependent on
soft money, which can be unpredictable. Postdocs and those who do pursue tenure
track positions need to publish and see no rewards in software development.
These challenges were all identified at WSSSPE1. What actions would we recommend
to improve things? Altmetrics and download statistics may slowly change the
system and improve a developer's ability to receive credits for time invested in
software development. NSF's recognition of scientific products including datasets,
software, and publications is also helping.

We asked ourselves, ``Are there examples of things that are changing
because of this and how can we build momentum?'' One example demonstrates change
over time. In 2007, nanoHUB listed the academic reward structure as a problem in
an EDUCAUSE report where the authors note, ``In the future, nanoHUB researchers
are hoping to change the research culture. While they recognize that young
faculty members are unlikely to get tenure based on their nanoHUB contributions,
they hope to encourage faculty to think beyond their own research needs to
consider publishing tutorials and other content in their fields on the
nanoHUB site.'' Fast forward seven years later where, in a 2014 iSGTW article,
quote nanoHUB Principal Investigator (PI) Gerhard Klimeck, ``A former student of
mine published eight tools on nanoHUB, serving over 6,000 people with his tools.
He then joined a university as a professor and introduced nanoHUB. Use of the
gateway from that university skyrocketed; he used nanoHUB in existing classes,
created new classes, and infused it in his research. Ultimately, the professor's
department head attributed his two-year rise to tenure with the reputation and
innovation he gained through nanoHUB.''

The group also discussed ``attribution trees,'' an idea put forward by Dan Katz
and Arfon Smith where a chain of attributions can, for example, give credit to
those developing libraries and building blocks that other pieces of software
use. The group considered potential journals and medium to push the application
of this idea. One participant noted that Dryad\footnote{The Dryad Digital
Repository, \url{http://datadryad.org}.} works with journals. If a paper is
accepted to one of those journals, the supporting data must be submitted to Dryad.
The group also discussed reproducibility as
a component of the journal review process and ``active papers,'' with immediate
links to the data and software.

We then discussed variations among scientific domain areas and wondered ``Are
some communities more or less open than others?'' To some members of the group,
biology seems to be more open, while physics less so. Some felt, with the more recent
development of bioinformatics as a field, there were fewer historical practices
to undo in biology. Physics has preprint philosophy to overcome. Environmental sciences
may be mixed. The group felt that the biomedical area, however, was very competitive
and closed. One participant mentioned blueprints for going open source (like
NWChem recently did) where authors outline how this helps, what you do and what
the next steps are.

We then moved beyond our training discussion to address funding that encourages
sustainable software -- funding of both people and projects that create a true
system to support sustainability. We called this ``institutionalized
serendipity.'' As science is increasingly reliant on software, one participant
observed that ``software development can have much broader impact than
publishing individual research, but it is not viewed that way.'' Because of this
centrality, one participant mentioned that training ought to be called Science
Carpentry rather than Software Carpentry. We felt we were beginning to see
changes in the research community as a result of the NSF's data management plan
requirements. PIs are thinking more about data and some university libraries are
offering data repositories. We wondered if a software management plan might be
effective. ``How might funding programs need to adapt to reward good software
development practices?'' we asked ourselves. We thought about best practices
such as version control, test harnesses, mailing lists, bug tracking, community
contribution, and reuse where appropriate. We thought about measuring success
through usage statistics (downloads, altmetrics). ``Should funders demand
reproducibility?'' we wondered. In order for results to be reproduced, software
would need to be carefully curated.

Again, our industry participants contributed unique viewpoints. Partnering with
industry was seen as one path to sustainability. The unique partnerships Kitware
engages in promote academic freedom while creating an income stream from certain
portions of the software. This type of approach to sustainable software frees
researchers from performing tasks that do not offer the rewards their
institutions value. We also discussed about successful models for industry
partnerships that preserve open science. Participants noted that there are some
NSF programs that prohibit partnerships with for-profit companies (but there are
other programs in which this is encouraged).

\subsubsection{Subgroup A Actions}

Three main actions were identified by Subgroup A that would be of interest to
participants and benefit the research community. These focused on the
desire to amplify the current community efforts in training and software
engineering by connecting the current training initiatives (e.g., Software and Data Carpentry,
rOpenSci, OSS) that are of different durations at appropriate stages.
Generally it was felt that a modicum of interaction between short term workshops
(SwC, DC), medium term trainings (e.g., OSS), and longer-term courses (e.g.,
BIDS Data Science) would benefit from coordinating curriculum, discussing and
aligning prerequisites, and coordinating timing of courses. A training
coordination effort would go a long way towards amplifying the value of the
individual efforts and make them all more effective.

\begin{itemize}
\item Action 1: Create a roadmap of research software training initiatives. Such
a roadmap would provide a taxonomy of training opportunities: what they deliver,
and what attendees need to know going in (prerequisites). It would also show
which recommended roadmap actions will have a clear and immediate payoff, and
which will have long-term payoffs. The training roadmap would emphasize time
savings and efficiency gains to be had from each training.

\item Action 2: Build a report card characterizing use of best practices in
scientific software. Generally, people felt that researchers would be very
willing to migrate practices if they could identify where they needed to
improve. Such a report card would ask simple questions to characterize use of
best practices in science software, and could be structured similarly to Joel
Spolsky's Software Maturity questionnaire (also known as
\href{http://www.joelonsoftware.com/articles/fog0000000043.html}{the Joel
Test}~\cite{joel-test}). The survey would create a report card that shows areas
where a project could improve, and then link those areas to specific training
offerings from the training taxonomy from Action 1.

\item Action 3: Create a science software review forum. It was generally
acknowledged that a little code review can have a tremendous impact on the
quality of software in a project, but that sites for science code review are
lacking. While people can ask questions on the Stack Exchange sites, they are
generally not open to questions of style, approach, or appropriateness, as they
try to avoid subjective commentary. Instead, we need a site where code gets
discussed/summarized/described (in small bites) by the science software
community. The target audience would be graduate and undergraduate students in
the sciences, and there would need to be mechanisms to keep the review positive
and constructive, and not get pedantic or judgmental. This could be tied to a
software registry (such as the nascent GeoSoft project), or to language
repositories like R's CRAN repository, and could lead to the report card
discussed in Action 2.
\end{itemize}

\subsubsection{Subgroup B Actions}

Subgroup B then focused on actions it could take. It discussed development of a
white paper that describes a matrix approach to training (multi-day, multi-week,
semester). The white paper might include a survey of existing techniques. There
are many, some dating back many years, for example the Interuniversity
Consortium for Political and Social Research (ICPSR) and various summer
institutes. The white paper could include a call for a comprehensive assessment
of these techniques. We need to think carefully about the right venue for such a
white paper, where it would have the most impact. The subgroup believes it would
need to approach editors directly to ascertain this.

The group felt that training in techniques that promote sustainability has a
range of benefits: career paths, educated reviewers, reproducible science, and
more. There is some information on how that has been approached and assessed,
but more is needed. The group felt that this training is undervalued and that it
is important to communicate the return on investment (ROI) -- both individual ROI
(skills that make scientists more effective and more marketable) and funder ROI
(better use of taxpayer funds, research more likely to be reproducible because
sustainable software exists, better trained reviewers).

\subsection{Papers}
The papers that were discussed in the Software Development Experiences Subgroup~A are:

\begin{itemize}
\item Michael R. Crusoe and C. Titus Brown. Channeling community contributions to
scientific Software: A Hackathon Experience~\cite{wssspe2_crusoe}

\item Marcus Hanwell, Patrick O'Leary, and Bob O'Bara. Sustainable software
ecosystems: Software Engineers, Domain Scientists, and Engineers Collaborating
for Science~\cite{wssspe2_hanwell}

\item W. Christopher Lenhardt, Stanley Ahalt, Matt Jones, J. Aukema, S. Hampton,
S. R. Hespanh, R. Idaszak, and M. Schildhauer. {ISEES-WSSI} Lessons for
Sustainable Science Software from an Early Career Training Institute on Open
Science Synthesis~\cite{wssspe2_lenhardt}

\item Jory Schossau and Greg Wilson. Which Sustainable Software Practices do
Scientists Find Most Useful?~\cite{wssspe2_schossau}

\end{itemize}

The papers that were discussed in the Software Development Experiences Subgroup~B are:

\begin{itemize}

\item Jordan Adams, Sai Nudurupati, Nicole Gasparini, Daniel Hobley, Eric
Hutton, Gregory Tucker, and Erkan Istanbulluoglu. Landlab: Sustainable Software
Development in Practice ~\cite{wssspe2_adams}

\item Alice Allen and Judy Schmidt. Looking before Leaping: Creating a Software
Registry~\cite{wssspe2_allen}

\item Carl Boettiger, Ted Hart, Scott Chamberlain, and Karthik Ram. Building
Software, Building Community: Lessons from the {ROpenSci} Project~\cite{wssspe2_boettiger}

\item Yolanda Gil, Eunyoung Moon, and James Howison. No Science Software Is an
Island: Collaborative Software Development Needs in
Geosciences~\cite{wssspe2_gil}

\item Ted Habermann, Andrew Collette, Steve Vincena, Werner Benger, Jay Jay
Billings, Matt Gerring, Konrad Hinsen, Pierre de Buyl, Mark K\"{o}nnecke, Filipe
Rnc Maia, and Suren Byna. The Hierarchical Data Format ({HDF}): A Foundation for
Sustainable Data and Software~\cite{wssspe2_habermann}

\item Eric Hutton, Mark Piper, Irina Overeem, Albert Kettner, and James Syvitski.
Building Sustainable Software - The {CSDMS} Approach~\cite{wssspe2_hutton}

\item James S. Spencer, Nicholas S. Blunt, William A. Vigor, Fionn D. Malone, W.
M. C. Foulkes, James J. Shepherd, and Alex J. W. Thom. The {H}ighly {A}ccurate
{N-DE}terminant ({HANDE}) Quantum {Monte Carlo} Project: Open-source Stochastic
Diagonalisation for Quantum Chemistry~\cite{wssspe2_spencer}
\end{itemize}

\section{Credit \& Incentives} \label{sec:credit}


This group, with just three papers but a large amount of interest and
participation from attendees, focused on the institutional, social, and cultural
mechanisms that encourage the creation and maintenance of shared software, in the
context of what now exists, what mechanisms are desired, and how
we might achieve them.

\subsection{Discussion and Actions}

In the first discussion session, this group decided to break into two smaller groups,
each independently working through the same general topic: credit and incentives.

\subsubsection{First breakout discussion: Group A}
The first sub-group discussed issues around the current system for credit and
incentives, which it called ``hacking the incentive structure.'' The
group considered four potential points of leverage:

First, that we currently have systems that collect information, and these could
be modified to collect different information, then map that information to
actions. We could initially build a proof-of-concept for a new use of a given
system, then determine what actions would be needed to make this use more
common.

Second that we could create entirely new systems, perhaps because the existing
systems are too tied to what they measure, and modifying them is not practical.

Third that we could change academic culture, rather than worrying about the
systems. This was mostly focused on citations, because they matter for hiring,
promotion, and tenure decisions. The group discussed how we could weigh the
citations within papers better than we now do. How could we identify the five
citations that really matter for a paper, distinguishing them from the related
works and general background that are also cited? Perhaps we could break these
out in the reference list, working with publishers to implement this. Or maybe
we could also break out categories of citations, such as the most important
software used, the previous publication that we are building from, the data that
we actually used, etc. The Moore Foundation's award in data science was given as
an example, asking proposers: What are the five canonical citations that are
most important to your work?~\cite{moore-canonical} This would be a way of
giving credit and assigning importance to these works, differently from how we
just count citations today. A possible action that the group discussed was
conducting a longitudinal study of most useful software, data, etc.\ in a
discipline.

Fourth, that we could change the ways funders make decisions, and use these
funding policies as incentives.

After this discussion, group A brainstormed about incentives,
with the following items suggested:
\begin{itemize}
\item running programming contests, creating bounties for contributing to open
source software, etc.
\item augment author lists to give credit to people who do not now get credit
(and making them machine readable)
\item developing a microcitation standard and mechanism (for both software and
data)
\item developing a well-defined standard for author contribution -- what level
of contribution rises to the level of authorship?
\item leveraging social media for citation and reviewing of content -- then
using social media to bring more people into the review process than is
traditional
\item determining where else software can be cited and recorded (e.g.,
acknowledgments sections of papers)
\item developing a taxonomy of contributors (e.g., Project
Credit~\cite{projectcredit}) tied to places that these metrics are already
stored (e.g., ORCID~\cite{orcid})
\item making metadata easier to add for software, creating incentive for
providing software metadata -- note that this cannot be centralized
\item creating something like the h-index that tenure committees can make use of
- simplify a way of measuring and documenting the overall credit given to an
individual over different projects
\item thinking about publishing software versus journal - software does not have to
be novel
\item determining guidelines for recommending software characteristics for
tenure - perhaps draft guidelines then get ACM or IEEE agreement.
\end{itemize}

\subsubsection{First breakout discussion: Group B}

This group started by discussing who should be incentivized, thinking of two
categories of people: those in science (who could be incentivized to do better,
more shareable, more sustainable work), and those in industry but interested in
science (who could be incentivized to contribute to science.) It was pointed out
that we are not yet clear enough on exactly what we want to incentivize,
suggesting that we need to have a clearer picture of ``good computational work''
and what sort of contributions are truly generative for science.

The discussion of incentivizing those in science acknowledged that the
publications system was far from perfect for incentivizing good software work.
Nonetheless, the discussion focused on bringing software people into publications.
There were two main suggestions. The first is to focus on end users of software
and encourage them to cite software the ``right'' way. James Howison suggested
that his research showed that few projects were making a formal request for
citation (but that authors weren't necessarily following those suggestions
anyway)~\cite{howison2015jasist}. He suggested making access to the software
conditional on a license that requires citation. Others found this ``too
confrontational'' and preferred to concentrate on making it easier to do the
right thing. The second was focusing on those leading software projects, and the
group was more enthusiastic about ``forcing'' PIs to include their ``software
people'' on publications, although there were few ideas on how exactly to do
this. Another technique mentioned was that when scientific software projects are
hosted in organizations like Apache, the scientific contributors can benefit
from building their reputations, perhaps yielding job offers that they can use
to negotiate better job and career packages.

The discussion on incentivizing those outside science focused on accessing the
well of affection that those working in software have for scientific research.
How can the interest and skills of this group be marshaled towards sustainable
contributions? There is evidence that the migration of scientific software
projects to the Apache Software Foundation (ASF) has created opportunities for those not
employed in the scientific center to contribute to projects initiated by
scientists (especially where there is cross-over with industry needs, such as
provenance and workflow).

The group also discussed developing ``software prizes'' arguing that while it is
hard to ``mint'' other new sources of reputation, prizes are possible without
getting too many others on board. The prize criteria can form a template for
describing what we mean by scientific contributions made through software,
particularly focusing on building active communities, not only writing great
code.

\subsubsection{Group merger and redivision}

After the first breakout session, the groups A and B came together and discussed
a rollup of the ideas from the subgroups at a high level:
\begin{itemize}
\item citation ecosystem -- traces of usage (metrics)
\item taxonomy of contributorship, understanding roles
\item prizes
\item new metrics (for people's activities in software)
\item guidelines for evaluating scientific contribution through software (perhaps
using new metrics).
\end{itemize}

In the remaining discussion sessions, the group chose to split into three
subgroups to discuss a version of these topics: {\em citation ecosystems}, {\em
taxonomy of contributorship/guidelines for software for tenure review}, and {\em
prizes}. The subgroups were asked to clearly identify
\begin{itemize}
\item the problem to be solved
\item steps towards a solution.
\end{itemize}

\subsubsection{Remaining breakout discussions: Citation ecosystems}


This group defined its goal as creating a low-barrier-of-entry method for
recording names and roles (and in a second phase, optionally including weights)
of contributors (coders and other intellectual contributions) to a software
package in a machine readable way (to be called a credit file), then encouraging
the scientific community to adopt this practice.

The following general points were initially discussed:
\begin{itemize}
\item The FLASH~\cite{flash} project was suggested as an example of how something
like this has been done.
\item This data could be a file that can be associated with a citation to the
software, either through use of a DOI (digital object identifier) for the credit file, or by uploading the
credit files as associated with the paper.
\item This idea could also be applied to data.
\item The credit file should be part of the metadata that are freely
available to those who have not paid for access to the journal, like citations
are now in most cases.
\item It was suggested that there should also be a \emph{separate} file to track
software dependencies.
\end{itemize}

The group came up with the following actions to be performed:
\begin{enumerate}
\item Build a tool that can automatically determine who the contributors are
(from a Git or other repository), then allows the user to manually edit the output to
add/remove people, define roles.
\item Work with repositories to encourage them to provide the information we
need based on what they already store.
\item Define what a citation file should look like and what it should be called.
\item Test adoption with a substantial scientific organization such as the Lawrence Berkeley National Laboratory (LBNL).
\item Create credit file for a set of software.
\item Build a validator (and perhaps a visualizer) for credit files.
\item Write a tool to collect files and visualize/output interconnections (which
software is used with which), based on an existing project.
\item When we (the group members) write papers, we should track the software
we use, and encourage the software developers to make their software citable
and create credit files.
\item Build a tool to export the credit file to BibTex and other citation styles.
\item Make sure the BibTex entries (exported from internal data) are somewhat
standardized so that they can be imported into papers.  Also make sure that
standard LaTeX style files understand and accept these entries.
\end{enumerate}

\subsubsection{Remaining breakout discussions: Taxonomy of contributorship/guidelines for software for tenure review}


At the start of the discussion, the breakout group brought forth the important
observation of the wide disparity in commonly accepted habits of publication in
different research fields. In domains which have, historically, relied on large
groups of researchers collaborating towards a common goal (e.g., high-energy
physics, astronomy), publications often have tens or even hundreds of co-authors
(with some papers in experimental particle physics having over 3000). In other
domains, the number of co-authors is typically much smaller with, in some
cases, even a preference for single-author papers. Similarly, the various
platforms for publication are valued differently in different domains. Most
commonly, publications in peer-reviewed scientific journals are regarded as the
most important and most impactful. However, in certain domains, especially in
Computer Science, many researchers typically regard conference proceedings as
their prime publication target. It is often suggested that this difference is
due to the rapid developments in information technology, a pace that cannot be
upheld by traditional peer-reviewed journals. Whatever the causes, any useful
taxonomy of contributorship or guideline for tenure review should take such
differences into account.

Despite these differences, and despite the fact that software often has taken
the role of a proper, albeit less tangible, scientific research instrument,
neither the software nor its creators are commonly credited as part of a
scientific publication. The group acknowledged the need for more recognition for
the creators of such software instruments, and indicated a number of possible
pathways. First and foremost, domain scientists must be made aware of the
important role of software, and include the developers as co-authors of papers.
A second approach is to fully embrace an open badging infrastructure (such as
Mozilla's Open Badges), where a {\em badge} is a free, transferrable,
evidence-based indicator of an accomplishment, skill, quality, or interest. A
third approach is for the scientific community to support the increasing
momentum of peer-reviewed journals specialized in the open source/open access
publication of scientific research software, such as Computer Physics
Communication, F1000 Research, Journal of Open Research Software, and SoftwareX.

Recognizing publication of research software as a proper scientific contribution
raises several important but currently unsolved questions, however. For example,
is the number of users of the software a relevant measure of impact? What
standards of coding quality must be followed in order to justify publication and
hence recognition? Should the release of a new version of the software be
eligible for a new publication; if so, under what conditions? And above all:
should software publications be valued in the same way as traditional scientific
publications? Or is there a need for new measures of productivity and impact?

In part, the answers will come from the scientific community at large, as a
natural consequence of growing awareness and mindset change. Some of the
answers, however, also should be based on decades of experience in (and
developing standards for) implementing, maintaining, refactoring, documenting,
testing, and deploying software instruments in scientific research. Care should
be taken, however, not to impose such standards for all domains in equal ways
right from the start. Forerunners should serve as an example, but should not
scare away domains that have based their progress on much less advanced methods
of {\em software carpentry}. Nevertheless, proper guidelines are needed, which
eventually should be followed across all domains. The group also recognized that
funding bodies, universities, and publishers eventually should demand that
research projects follow such guidelines, and to implement a proper software
sustainability plan.

To enable a form of standardized crediting for developers of research software,
the group proposed to work towards a taxonomy for software-based
contributorship. The taxonomy should be derived from, or extend, existing
taxonomies for research impact and contributorship such as defined by CASRAI (in
particular based on the Wellcome-Harvard contributorship
taxonomy\footnote{Project CRediT, \url{http://credit.casrai.org/}}, VIVO, or
ISNI. An interesting measure of impact raised by the group was the {\em
betweenness centrality}, an indicator of a person's centrality (and hence,
importance) in a scientific collaboration. It is expected that developers of
research software often play such a central role. 

The group defined the following actions to be performed:
\begin{enumerate}
\item Investigate existing taxonomies for roles and contributorships.
\item Investigate prototype badging initiatives.
\item Investigate journals focusing on publishing peer-reviewed research
software.
\item Investigate guidelines and checklists of best practices.
\item Communicate the results of the above investigations to the WSSSPE
community and decision-making bodies (funders, publishers, universities,
and tenure committee representatives).
\item Ensure engagement of the broader research community in this discussion.
\end{enumerate}

\subsubsection{Remaining breakout discussions: Prizes}\label{sec:prizes}

This group discussed the idea of prizes. Prizes are expected to reframe software
as ``instrument building'' but will prizes be good or bad, and how can we make
sure there are no negative affects and the process cannot be gamed?

Prizes in different categories were discussed (like Academy Awards), for
example: best contribution (non-founder), broadest diversity of contributions,
best tutorials or documentation, best leadership transition (award ex-leader and
new leader), best generalization (taking something that was limited and making
it more general), and best mentorship of contributors (bringing others into the
community).

Other (non-prize) forms of incentives are
\begin{itemize}
\item Converting reputation by joining the ASF, Google, etc.
\item Inviting ASF and open source people to contribute to scientific code (at
least in areas where there is overlapping interest).
\item Template for assessing scientific contributions made through software.
\end{itemize}

The group suggested that to determine who should give out prizes, perhaps we
should find those who we think would be awarded prizes, then ask them who they
would want to receive a prize from, and subsequently contact those organizations to see if they
are willing to be involved in the process.

One of the group's ideas was to create a funding program for disciplines or
other organizations to create a prize program. We would provide a framework and
a set of requirements, for example: awards to individuals; must award in 5 or 6
areas; must have a jury that includes senior/junior
domain experts and technologists (and should have objective criteria); must have the recipients awarded at a
relevant event; and must provide citations that explain why the prizes should be
awarded. Different organizations could then decide to sign up to the framework
and give out awards under this general brand. However, there was a concern that
having many organizations award prizes may reduce the impact of the prizes.

Possible groups that might give out awards, either under our framework, or more
generally, are AAAS, Nature Publishing Group, ACM, IEEE, Astro, Ecology Society
of America, etc. Perhaps this could be a joint technology/science partnership,
for example, the [Apache $|$ Mozilla]-[AAAS $|$ disciplinary society] prize.

Some potential criteria for prizes the group suggested are:
community engagement, helping out others;
number of unique contributors;
adding new pieces of functionality to software;
integrating software into broader ecosystem; championing broad principles of
sustainability, open science, open source, etc.;
improving accessibility to software, to scientific software (perhaps championing
inclusiveness or making software accessible?);
documentation;
tutorials;
commits/patching;
leadership transition; and
best contribution by a non-founder.

The group discussed if there should be different criteria for ``established''
members of the community versus junior members, and if prizes should be
restricted to junior members, but left these as open questions.

A point the group considered important is that we do not want to give prizes just
to reward people who are really good at this one thing, but rather we want to
reward people who are building the culture we want as scientists.

\subsection{Papers}
The papers that were discussed in the Citation \& Incentives group are:
\begin{itemize}
\item James Howison. Retract Bit-rotten Publications: Aligning Incentives for
Sustaining Scientific Software~\cite{wssspe2_howison}

\item Daniel S. Katz and Arfon M. Smith. Implementing Transitive Credit with
{JSON-LD}~\cite{wssspe2_katz}

\item Ian Kelley. Publish or Perish: The Credit Deficit to Making Software and
Generating Data~\cite{wssspe2_kelley}
\end{itemize}

\section{Reproducibility \& Reuse \& Sharing} \label{sec:reproduce}


This group discussed five papers with a wide variety ideas of how to support
reproducibility and reuse. It focused on identifying concrete practices that the attendees
could work together on, which would have a positive effect on the community.

\subsection{Discussion and Actions}

In the first discussion session, this group broke up into two smaller groups
working on different topics: Reproducibility, and Reuse and Sharing. In the
second discussion session, three subgroups worked on creating specific pieces of
guidance.

\subsubsection{First breakout discussion: Reproducibility group}

The first group discussed ways in which reproducibility of papers could be
improved. A consensus surrounded the provision of examples: demonstrating to
others how to achieve reproducibility. Major public funding investments go into
research that heavily relies on reproducible software, hence the lack thereof
is raising major concerns.

Two main avenues could be used to implement policy that would improve
reproducibility and drive top-down culture change:
\begin{itemize}
\item Funders can aim to get more software expertise on to funding review panels
and provide more guidance on what is required of software related to research
(cf.\ data management plans)

\item Journals can define and publish journal policies to improve
reproducibility, and reviewers can insist that authors provide sufficient
information and access to data and software to allow them reproduce the results
in the paper. Stronger policies even for some high-impact journals have recently
come into place, for example Nature Publishing journals.
\end{itemize}

If journals do enforce greater reproducibility constraints, it is important to
lower the barriers to reviewers attempting to verifying the correctness of the
software used to generate the results. A major issue is that a lot of software
only builds on certain systems. Should journals provide more tools/support for
reviewers, and if so, what is it? An idea that came from the groups was to
define a set of support services that should be available to software paper
reviewers. Another was to provide the ability for reviewers to flag the
requirement for a `software verification,' similar to the ability to flag that a
paper needs to be seen by a statistician.

Other discussion focused on ways in which researchers themselves could improve
the reproducibility of research. One way would be to establish tracks at
conferences that subject papers to reproduction, which for those that pass would
lend them an additional badge (similar to OOPSLA). Another is simply to get more
people to use your software: for instance by outreach to high school students --
can your software be used by them? Finally, there is a role for
community-curated benchmarks to validate the performance and capabilities of
tools.

\subsubsection{First breakout discussion: Reuse and Sharing group}

The second group discussed ways in which software could be more easily reused
and shared. From the perspective of both user and developer, any solutions must
be 1) easy; 2) cheap; 3) not too time-consuming.

Reuse and sharing were considered to be distinct but linked. In many cases,
pre-existing software does not exist, so new software is written but even then
it is not shared afterwards. The principal barriers to reuse are the difficulty
of finding out what software is available, understanding if it is usable, and
then of installing and running software if it is located.

Discovery of relevant software is still a fundamental issue: we need standard
vocabularies and metadata, better tools, and approaches that are sustainable.
Publications are an easy entry point for locating suitable software, but should
the publishers lead the way, or is this the responsibility of ``the community''?
Some communities have had significant initiatives to improve software discovery,
e.g., the NIH Software Discovery Index\footnote{NIH Software Discovery Index:
\url{http://softwarediscoveryindex.org/}}. Likewise, there were examples of
journals which had made software more discoverable: ACM Transactions on Software
offers a reproducibility review; the Journal of Biostatistics has an opt-in to
provide code and a certification mark if it can be run; the Journal of Open
Research Software requires software to be deposited in suitable repositories and
referenced.

An issue around the sustainability of software catalogues is that their
usefulness often depends on the domain. For instance, in the biosciences, there
is more homogeneous data and standard shared code. In areas like ecology,
code is often very specific to a problem, meaning that the level of re-use might
be at a general statistical level of abstraction, but then every research
use is highly customized.

Finally, it was clear from the discussion that there were many ways in which
software could be made available in more reusable ways than just a tarball
sitting on a personal website. Using code repositories like
GitHub\footnote{GitHub: \url{http://github.com/}} gets you an archived, shared
platform and improve the reusability of your software incrementally, for
instance by adding a license or by archiving (with a DOI) in
Figshare\footnote{FigShare: \url{http://figshare.com/}} or
Zenodo\footnote{Zenodo: \url{http://zenodo.org/}}. Docker\footnote{Docker:
\url{htp://www.docker.com/}} might be a solution to the issue of dependencies,
to allow binaries and libraries to be bundled in a more lightweight fashion than
a virtual machine image.

The key enabler for reuse and sharing was to get domain scientists more
effectively connected with programmers/analysts. Both have skills and experience
which is necessary to make the right decisions for improving the reusability and
discoverability of software, and to apply community pressure to change practice.

\subsubsection{Second breakout discussion: Categorization of software journals}

This group aimed to come up with a categorization of journals which published
software papers. Starting from the list of
journals\footnote{\url{http://www.software.ac.uk/resources/guides/which-journals-should-i-publish-my-software}}
maintained by the Software Sustainability Institute, the group chose seven
journals and looked at their advice to authors and reviewers. These were the
Journal of Open Research
Software\footnote{\url{http://openresearchsoftware.metajnl.com/about/editorialPolicies\#peerReviewProcess}},
PLoS ONE\footnote{\url{http://www.plosone.org/static/guidelines\#software}},
Journal of Statistical
Software\footnote{\url{http://www.jstatsoft.org/instructions}}. Methods in
Ecology and
Evolution\footnote{\url{http://www.methodsinecologyandevolution.org/view/0/authorGuidelines.html}},
Transactions of Mathematical
Software\footnote{\url{http://toms.acm.org/Authors.html}},
GigaScience\footnote{\url{http://www.gigasciencejournal.com/about}}, and PLoS
Computational
Biology\footnote{\url{http://www.ploscompbiol.org/static/guidelines\#software}}.
From these a set of common categories were synthesized, against which all
journals could be compared:

\begin{itemize}
\item Journal Policies
\begin{itemize}
	\item Accessibility of papers
	\begin{itemize}
		\item Open access
		\item ``Freely'' available
	\end{itemize}
	\item Repositories
	\begin{itemize}
		\item Provides suggestions for recommended repositories
		\item Provides its own repository
	\end{itemize}
	\item Review
	\begin{itemize}
		\item Reviewing software is mandatory
		\begin{itemize}
			\item Must check that software runs
			\item Must check quality of code
			\item Must check performance of code if paper makes claims on relative performance
		\end{itemize}
	\end{itemize}	
	\item Supporting data
	\begin{itemize}
		\item Must be publicly available
		\item Must be in an open access repository
		\item must have a DOI
	\end{itemize}
	\item Article processing charge (APC)
	\begin{itemize}
		\item APC charge is transparent
		\item APC waiver program
	\end{itemize}
\end{itemize}

\item Paper Policies
\begin{itemize}
	\item Required sections
	\item Keywords
	\begin{itemize}
		\item Paper provides keywords to help describe software
	\end{itemize}
	\item Papers can be updated when new releases of software are made
	\begin{itemize}
		\item At no extra cost/at significantly reduced cost
	\end{itemize}
\end{itemize}

\item Software Policies
\begin{itemize}
	\item Software must have a license
	\begin{itemize}
		\item Software must have an open source license
	\end{itemize}
	\item Availability
	\begin{itemize}
		\item Software must be openly available and accessible
	\end{itemize}
	\item Deposit policies
	\begin{itemize}
		\item Software should be in a public repository
		\begin{itemize}
			\item Of particular stature/with a preservation plan
		\end{itemize}
		\item Software should have a permanent identifier
		\item Software deposit doesn't count as a prior publication
	\end{itemize}
	\item Runnability and dependencies
	\begin{itemize}
		\item Provide documentation to understand how to run
		\item Provide sample data
		\item Provide dependency information
	\end{itemize}
\end{itemize}
\end{itemize}

The follow-up actions to this work are to use this set of categories on all the
journals in the list, refining the categories if necessary, then identify if any
of the categories are seen to be more useful to promote reproducibility, reuse
and sharing.

\subsubsection{Second breakout discussion: What should journals provide reviewers of software papers?}

This group discussed whether they could come up with a list of things that a
journal should provide its reviewers to make it easier to review software
submitted for publication as a ``Software Paper.''

Journals should provide guidelines about what to consider when reviewing a
software submission. A good example for a relatively comprehensive guidelines
for reviewers (and thus in turn authors) are those of
JORS\footnote{\url{http://openresearchsoftware.metajnl.com/about/editorialPolicies}}.
Journals might also learn from organizations such as the ASF as to
what is good practice for software submissions. Guidance is needed on what
constitutes an incremental improvement that is significant enough to qualify for
publication, otherwise this assessment can be very subjective.

Journals should provide mechanisms to enable and track communication between
reviewers and authors. For anonymous peer review journals, anonymity of
reviewers should be maintained. If communication is necessary, that may mean that software is not
that well documented. Journals should also provide a set of simple metrics for
software evaluation that reviewers can use for ratings, similar to Consumer
Reports.

Journals should provide guidelines about requirements for documentation of code:
both in-lined in code, and manuals/web pages, etc. Journal editors could provide
documentation checks before it goes out for review. This should include a
requirement for good Use Cases specified for the software, with references to
executable test cases that demonstrate each use-case is met (at least in the
form of the test case).

Journals should support mechanisms to run software. Sometimes this may be very
hard to accomplish, despite best efforts. Journals could provide an execution
environment for any software submitted for review, perhaps via a Docker
container or a virtual machine. If not, instructions must be adequate to compile
and execute the software; to interpret the results (output files, formats,
etc.); and the full source code must be accessible to the reviewer. Mechanisms
to quantify what has changed compared to a previously published version would
assist version comparison. However, this cannot simply take lines of code into
account. For example, a speedup of an algorithm may not result in a huge code
difference, but may nonetheless provide a breakthrough. Journals should require
software submissions to also provide `test materials' -- sample data, parameters
to validate that code is working as intended. In certain cases, well-selected
benchmark datasets may be required to assess performance and accuracy.

Journals should ensure minimal metadata are provided, similarly as is already
the case for certain kinds of data (though the latter is often enforced by data
repositories) -- Dublin core-ish (creator, owner), and more specific (platform
and compiler dependencies, sample benchmarks of performance, etc.). Journals
should provide guidance or constraints as to software licensing conditions.

The follow-up actions to this work are to work with journals and reviewers to
identify whether any of these suggestions can be easily provided, perhaps for a
range of journals.

\subsubsection{Second breakout discussion: Reproducibility Meta-track toolkit}

This group worked on defining a ``toolkit'' for running a reproducibility
meta-track at a conference. They decided to take the work done during the
workshop and publish it as a paper.

\subsection{Papers}
The papers that were discussed in the Reproducibility \& Reuse \& Sharing group are:
\begin{itemize}
\item Jakob Blomer, Dario Berzano, Predrag Buncic, Ioannis Charalampidis,
Gerardo Ganis, George Lestaris and Ren\'{e} Meusel. The Need for a Versioned
Data Analysis Software Environment~\cite{wssspe2_blomer}

\item Ryan Chamberlain and Jennifer Schommer. Using {Docker} to Support
Reproducible Research~\cite{wssspe2_chamberlain}

\item Neil Chue Hong. Minimal Information for Reusable Scientific
Software~\cite{wssspe2_chue_hong}

\item Tom Crick, Benjamin A. Hall, and Samin Ishtiaq. ``Can I Implement your
Algorithm?'': A Model for Reproducible Research Software~\cite{wssspe2_crick}

\item Bryan Marker, Don Batory, Field G. Van Zee, and Robert van de Geijn. Making
Scientific Computing Libraries Forward Compatible~\cite{wssspe2_marker}

\item Stephen Piccolo. Building Portable Analytical Environments to Improve
Sustainability of Computational-Analysis Pipelines in the
Sciences~\cite{wssspe2_piccolo}
\end{itemize}

\section{Code Testing \& Code Review} \label{sec:code_testing}


This group began by recognizing that they knew what the problems of code testing and code
review are, so the purpose of the group was really to think about
concrete things that people in the group would be willing to commit to, in order
to improve the sustainability of software.

\subsection{Discussion and Actions}

The consensus of the group was that small, concrete actions would have the
greatest chance of proper followthrough by members of the group. An example of
this approach was the Architecture of Open Source Applications book~\cite{aosa},
containing contributed chapters on open source software. The group contemplated
creating an analogous book or web resource for testing of scientific software.

A key point of discussion was that there are two major challenges to software
testing in science. The first is convincing developers to incorporate testing.
This challenge is both social and technical. We need to communicate the value of
testing and also teach developers how to choose and use testing frameworks. The
second challenge---and the one that the group felt was more difficult---is
choosing appropriate tests for scientific software. This is the difference
between learning how to use, say, Python \texttt{unittest} and knowing how to test
that one's code correctly implements, for example, a Lattice-Boltzmann model in
computational fluid dynamics.

\subsubsection{Choosing and implementing testing frameworks} There was general
consensus that once a developer has been exposed to a testing framework in a
nontrivial fashion, he/she will subsequently insist upon using a framework for
further work. The acknowledged challenge was how to create that crucial first
exposure to such frameworks.

Within the group, there were several different paths by which participants had their 
first exposure to code testing:
\begin{itemize}

\item Formal tutorials at conferences (for example, a test driven development
(TDD) at a Software Developers Best Practices conference): These often require
self-learning following a tutorial, but it can be hard to find a tutorial that
matches the programming language and/or the domain of the participant.

\item As a way to be confident in other software: When breaking out pieces of
software from a larger application, a developer wants to ensure the software
works as expected. These types of tests may be ephemeral though, living long
enough to give the author of the test confidence in the code, but not handed on
and made available to next user. These experiences can lead to more systematic
testing.

\item Through experience with coding: As you write more software, your
confidence in code is reduced and the importance of testing becomes clearer.
Attendees joked that they started testing ``when their skepticism/guilt became
larger than their arrogance.''

\item From other projects: When building on top of projects with good testing
frameworks, you realize that you should also adopt those good software
engineering habits.

\end{itemize}

The two non-self-taught paths were direct learning and indirect through other
software (learning by example). The group recognized that teaching about testing
needs to start early in projects and careers. We need to teach people how to
test short bits of code, rather than waiting until they have thousands of lines
of code and being surprised when they don't write tests. The group generated a
set of practical suggestions for improving adoption of code testing.
Established software projects can encourage the latter through a simple rule of
accepting only software patches with associated tests and through good
documentation of testing practices and requirements.

Simpler, standard ways of setting up appropriate testing infrastructure will be
important for adoption by scientists. Jenkins\footnote{\url{https://jenkins-ci.org/}} 
was suggested as a common open-source
solution, but the initial configuration was considered to be challenging for
typical scientists.

Teach testing by beginning with a (smallish) piece of code that lacks
appropriate tests and develop an exercise that involves refactoring the code
into a more presentable form through creating/deriving unit tests. Students
would extract unit-tests through reverse-engineering and/or questioning a
partner who is a developer/expert.

This led into a more general discussion about whether programming courses
should be part of the standard curriculum for science students and the
challenges of fitting additional material into university curricula. This is the
reasons that short workshops such as Software Carpentry and others exist.

Ironically, Software Carpentry no longer teaches testing~\cite{SCtesting}. This
boils down to two issues. First, that scientific computing doesn't (yet) have
the cultural norms for error bars that experimental sciences have, and second,
that there is a breathtaking diversity of scientific code; scientific research
is highly specialized, which means that the tests scientists write are much less
transferable or reusable than those found in other fields.

\subsubsection{Writing tests for scientific software}
The second major discussion thread in the group addressed the issues identified
by Software Carpentry -- the difficulties in writing tests appropriate for
scientific software: 
What tests are
appropriate to ensure that an complex method is working correctly? When is the
result of a numerical computation `close enough' to pass a test?
 
Several members of the group indicated that while they valued the concept of
software testing in theory, they were unaware of how to test certain kinds of
software. Or, more importantly, they lacked relevant examples of software
testing that would be suitable for new members of their teams. As an example, the
group asked if it could produce a 100-line example of how to test neutron
transport (or other specific scientific examples) targeted at a sophomore? This
would be a demonstration of ``How do I do this right?''

Another question was ``How do I test so that I know that the code is not the
source of my problems?'' A common challenge is that of stripping things down to
an appropriate level for tests. The group recognized that the difficulty is
different for so-called software infrastructure than for numerical/scientific
layers of software applications.
%
%
%
%
There is some discussion of the issue in the computer science
literature~\cite{hook1,hook2} but less in domain-specific journals. Some lessons
on multiphysics software verification can be found in a recent
paper~\cite{dubey-FLASH}.

Two important pieces of the barrier are a) picking/implementing testing
frameworks (a technical barrier) and b) deciding what are the actual tests that
I need to write? A possible method to solve the latter is to ask students ``how
is what you are doing different from typing random keys on the keyboard?'' and
then, turn their answer into the concept of tests. That is to say, expand the
process of creating code from one that is solely about ``how?'' into one that
also includes ``why?''

Another point of discussion is that many scientists are not aware that ``testing
the code'' and ``testing the science'' are distinct issues. A related question
that came up is ``How do you know that your method produces a result that is
`close enough'?''

The group then decided that a useful action would be to create a set of
scientific codes with associated tests. They developed the following basic
structure for these examples and sketched out six specific examples:
\begin{enumerate}

\item A paragraph or two explaining the scientific problem the code addresses

\item The size of the simplest piece of relevant, interesting code (i.e., an
estimate of lines of code)

\item A point-form list of the test cases you would use.

\end{enumerate}

The follow-up actions from this work include creating a compilation of testing
examples in scientific software. Some of the examples from the workshop have
become the starting seed of a collaborative book
(https://github.com/swcarpentry/close-enough-for-scientific-work) in which
scientists provide concrete examples of testing scientific software. The goal is
a set of testing examples, aimed at sophomores in science and engineering, that
cover a broad range of domains and problems and could be easily incorporated
into other workshops or courses.
 

\subsection{Papers}

The papers that were discussed in the Code Testing \& Code Review group are:
\begin{itemize}
\item Thomas Clune, Michael Rilee, and Damian Rouson. Testing as an Essential
Process for Developing and Maintaining Scientific Software~\cite{wssspe2_clune}

\item Marian Petre and Greg Wilson. Code Review for and by
Scientists~\cite{wssspe2_petre}

\item Andrew E. Slaughter, Derek R. Gaston, John Peterson, Cody J. Permann,
David Andrs, and Jason M. Miller. Continuous Integration for Concurrent {MOOSE}
Framework and Application Development on {GitHub}~\cite{wssspe2_slaughter}
\end{itemize}

\section{Conclusions} \label{sec:conclusions}

The WSSSPE2 workshop continued our experiment from WSSSPE1 in how we can
collaboratively build a workshop agenda, and we began a new experiment in
how to build a series of workshops into an ongoing community activity.

The differences in workshop organization in WSSSPE2 from WSSSPE1
are in using an existing service (EasyChair) to handle submissions and reviews,
rather than an ad hoc process, and using an existing service (Well Sorted) to
allow collaborative grouping of papers into themes by all authors, reviewers,
and the community, rather than this being done in an ad hoc manner by the
organizers alone.

The fact remains that contributors also want to get credit for their
participation in the process. And the workshop organizers will want to make
sure that the workshop content and their efforts are recorded. Ideally, there
would be a service that would index the contributions to the
workshop, serving the authors, the organizers, and the larger community. 
Since there still isn't such a service today, the workshop organizers are
writing this initial report and making use of arXiv as a partial solution to
provide a record of the workshop.

\begin{table*}[t]
\centering
\caption{Top tweets tagged \#WSSSPE on Nov 16, 2014.}\label{tab:tweets}
  \begin{scriptsize}
  \begin{tabular}{l|l|r|r}
 \hline
    Author  &   Tweet  & Retweets &  Favorites
\\ \hline
%
Neil P Chue Hong   &  Here's @SoftwareSaved guidance on Writing and using a software & 7 & 4
\\     &  management plan used by EPSRC software grants    &    &
\\     &  \url{http://www.software.ac.uk/resources/guides/software-management-plans}  &    &
\\  Neil P Chue Hong  &  @jameshowison as well as software plans   & 4 & 7
\\   &   \url{http://www.software.ac.uk/resources/guides/software-management-plans} &    &
\\   &   we provide a software evaluation tool:   &    &
\\   &   \url{http://www.software.ac.uk/online-sustainability-evaluation}  &    &
\\ Tom Crick  & $56\%$ of UK researchers develop their own software $\rightarrow  140,000$   &  14 & 8
\\ & UK researchers write research software w$/$out any formal training &    &
\\ Karthik Ram  &  OH: ``Institutionalize metadata before metadata institutionalizes you'' & 8 & 6
\\ Josh Greenberg  &  @jameshowison: ``1. retract any paper with bitrotten dependencies'' *mic drop* & 13 & 8
\\   &   ``2. add anyone who fixes bitrot as an author'' *mic drop*  &    &
\\ Ethan White &  ``@rOpenSci is all about community... our measures of success  & 9 & 3
\\ &   [include] how many faces are up on our community page''   &    &
\\ Ethan White & Daniel Katz talking about implementing transitive credit for  & 9 & 7
\\ &  software \url{http://arxiv.org/abs/1407.5117}  Work with @arfon  &    &
\\ Kaitlin Thaney  & Great point by @tracykteal about planning for ``end of life'' with scientific & 4 & 8
\\ &  software projects and sustainability.  & &
%
%
\\ Aleksandra Pawlik  & Lack of training as one of the main barriers for sustainable software & 10 & 4
\\ &    at @Supercomputing. @swcarpentry @datacarpentry can fix that!  & &
\\ Kaitlin Thaney  & My slides from this morning's keynote at & 11 & 12
\\ &  WSSSPE on Designing for Truth, Scale $+$ Sustainability:  & &
\\ &  \href{http://www.slideshare.net/kaythaney/designing-for-truth-scale-and-sustainability-wssspe2-keynote}{http://www.slideshare.net/kaythaney/}     & &
\\ &  \href{http://www.slideshare.net/kaythaney/designing-for-truth-scale-and-sustainability-wssspe2-keynote}{designing-for-truth-scale-and-sustainability-wssspe2-keynote} & &
\\ Neil P Chue Hong & @kaythaney shout out for @swcarpentry @datacarpentry & 9 & 4
\\ &  @rOpenSci @stilettofiend around open training activities for sustainability  & &
%
%
\\ Neil P Chue Hong & For those interested in Github - Figshare/Zenodo integration, & 5 & 12
\\ & but want SWORD/DSpace/Fedora/ePrints see:  & &
\\ & \url{http://blog.stuartlewis.com/2014/09/09/github-to-repository-deposit/}  & &
\\ Hilmar Lapp & Re: adopting the unix philosophy, consider signing the Small Tools in & 7 & 6
\\ & Bioinformatics Manifesto: \url{https://github.com/pjotrp/bioinformatics}   &
\\Andre Luckow &  ``Traditions last not because they are excellent, & 12 & 3
\\ & but because influential people are averse to change...''  C. Sunstein     & &
\\ Tom Crick &  ``Can I Implement Your Algorithm?'':  A Model for  Reproducible & 9 & 8
\\   &  Research Software \url{http://arxiv.org/abs/1407.5981}  & &
\\Mozilla Science Lab & At a loose end this Sunday? Care about reproducibility, software  $+$ & 10 & 5
\\ &  \#openscience? Follow the    \#WSSSPE hashtag for more, live from New Orleans.  &  &
\\Kaitlin Thaney & I'm in New Orleans at \#WSSSPE , speaking at 9:50 ET on  scientific software & 9 & 9
\\ &   $+$ sustainability. Tune in! Live stream: \url{http://ustre.am/17ddh} & &
\\   Daniel S. Katz & \#WSSSPE Agenda (Sunday):  & 10 & 1
\\ & \url{http://wssspe.researchcomputing.org.uk/wssspe2/agenda/}   &  &
\\ & URL for live stream of keynotes \& lightning talks: \url{http://ustre.am/17ddh}   &  &
\\ \hline
    \end{tabular}
    \end{scriptsize}
\end{table*}

WSSSPE actively used the online social network Twitter, with hashtag
``\#WSSSPE''. There were substantially more tweets (messages) during the days of
the workshops WSSSPE2, WSSSPE1.1, and WSSSPE1. Out of about 670 tweets as of Apr
18, 2015, more than 225 were about WSSSPE2 and about 180 were posted during the
day of the workshop. Some of the main points and highlights in the meeting are
shown in Table~\ref{tab:tweets}, which summarizes the top \#WSSSPE tweets from
the day of workshop, selected by the metrics that number of retweets or
favorites larger than five and the sum of two measures greater than ten.

In terms of building community activities, we wanted to focus primarily on
working groups, which we were able to do, as discussed above, but we
also wanted to make sure that attendees felt they had a chance to get their
ideas across to the whole group, which was the purpose of the lightning talks.
Overall, this seemed to be successful at the time, in terms of both the lightning
talks and the breakout groups, and the discussion of sustainability also led
to interesting and useful results. However, the challenge that we have discovered
since WSSSPE2 is that it is very hard to continue the breakout groups'
activities.  The WSSSPE2 participants were willing to dedicate their time to
the groups while they were at the meeting, but afterwards, they have gone
back to their (paid) jobs.  We need to determine how to tie the WSSSPE
breakout activities to people's jobs, so that they feel that continuing them
is a higher priority than it is now, perhaps through funding the participants,
or through funding coordinators for each activity, or perhaps by getting
the workshop participants to agree to a specific schedule of activities during the
workshop.

\section*{Acknowledgments} \label{sec:acks}

Work by Katz was supported by the National Science Foundation while working at
the Foundation. Any opinion, finding, and conclusions or recommendations
expressed in this material are those of the author(s) and do not necessarily
reflect the views of the National Science Foundation.


\appendix
\section{Attendees}  \label{sec:attendees}
The following is a partial list of workshop attendees who registered on the
collaborative notes document~\cite{WSSSPE2-google-notes} that was used
for shared note-taking at the meeting, or who participated in a breakout groups
and were noted in that group's notes.

{\small
\begin{longtable}{ll}
   Jordan Adams          &  Tulane University
\\ Alice Allen           &  Astrophysics Source Code Library (ASCL)
\\ Gabrielle Allen       & University of Illinois Urbana-Champaign
\\ Pierre-Yves Aquilanti &  TOTAL E\&P R\&T USA
\\ Wolfgang Bangerth & Texas A\&M University
\\ David Bernholdt       &  Oak Ridge National Laboratory
\\ Jakob Blomer
\\ Carl Boettiger        &  University of California Santa Cruz \& rOpenSci
\\ Chris Bogart          &  ISR/CMU
\\ Steven R. Brandt      &  Louisiana State University
\\ Neil Chue Hong        &  Software Sustainability Institute \& University of Edinburgh
\\ Tom Clune             &  NASA GSFC
\\ John W. Cobb
\\ Dirk Colbry           &  Michigan State University
\\ Karen Cranston        &  NESCent
\\ Tom Crick             &  Cardiff Metropolitan University, UK
\\ Ethan Davis           &  UCAR Unidata
\\ Robert R Downs        &  CIESIN, Columbia University
\\ Anshu Dubey           &  Lawrence Berkeley National Laboratory
\\ Nicole Gasparini      &  Tulane University, New Orleans
\\ Yolanda Gil           &  Information Sciences Institute, University of Southern California
\\ Kurt Glaesemann       &  Pacific northwest national lab
\\ Sol Greenspan         &  National Science Foundation
\\ Ted Habermann         &  The HDF Group
\\ Marcus D. Hanwell     &  Kitware
\\ Sarah Harris          &  University of Leeds
\\ David Henty           &  EPCC, The University of Edinburgh
\\ James Howison         &  University of Texas
\\ Maxime Hughes
\\ Eric Hutton           &  University of Colorado
\\ Ray Idaszak           &  RENCI/UNC
\\ Samin Ishtiaq         &  Microsoft Research Cambridge, UK
\\ Matt Jones            &  University of California Santa Barbara
\\ Nick Jones            &  New Zealand eScience Infrastructure, University of Auckland
\\ Daniel S. Katz        &  University of Chicago \& Argonne National Laboratory
\\ Ian Kelley
\\ Hilmar Lapp           &  National Evolutionary Synthesis Center (NESCent) \& Duke University
\\ Christopher Lenhardt
\\ Richard Littauer      &  University of Saarland
\\ Frank L\"{o}ffler     &  Louisiana State University
\\ Andre Luckow          &  Rutgers
\\ Berkin Malkoc         &  Istanbul Technical University
\\ Kyle Marcus           &  University at Buffalo
\\ Bryan Marker          &  The University of Texas at Austin
\\ Suresh Marru          &  Indiana University
\\ Robert H. McDonald    &  Data to Insight Center/Libraries, Indiana University
\\ Rupert Nash
\\ Andy Nutter-Upham     &  Whitehead Institute
\\ Abani Patra           &  University at Buffalo
\\ Aleksandra Pawlik     &  Software Sustainability Institute
\\ Cody J. Permann       &  Idaho National Laboratory
\\ John W. Peterson      &  Idaho National Laboratory
\\ Benjamin Pharr        &  University of Mississippi
\\ Stephen Piccolo       &  Brigham Young University, Utah
\\ Marlon Pierce         &  Indiana University
\\ Ray Plante            &  NCSA, University of Illinois Urbana-Champaign
\\ Sushil Prasad         &  Georgia State University, Atlanta
\\ Karthik Ram           &  Berkeley Institute for Data Science, University of California Berkeley \& rOpenSci
\\ Mike Rilee            &  NASA/GSFC \& Rilee Systems Technologies
\\ Erin Robinson         &  Foundation for Earth Science
\\ Mark Schildhauer      &  NCEAS, Univ. California, Santa Barbara
\\ Jory Schossau         &  Michigan State University
\\ Frank Seinstra        &  Netherlands eScience Center
\\ James Shepherd        &  Rice University
\\ Justin Shi
\\ Ardita Shkurti        &  University of Nottingham
\\ Alan Simpson          &  EPCC, The University of Edinburgh
\\ Carol Song            &  Purdue University
\\ James Spencer         &  Imperial College London
\\ Tracy Teal            &  Data Carpentry
\\ Kaitlin Thaney        &  Mozilla Science Lab
\\ Matt Turk             &  NCSA, University of Illinois Urbana-Champaign
\\ Colin C. Venters      &  University of Huddersfield
\\ Nathan Weeks
\\ Ethan White           &  University of Florida/Utah State University
\\ Nancy Wilkins-Diehr   &  San Diego Supercomputer Center, University of California San Diego
\\ Greg Wilson           &  Software Carpentry
\end{longtable}
}

\bibliographystyle{vancouver}

\bibliography{wssspe}
\end{document}